# Comments to "A non-thermal laser-driven mixed fuel nuclear fusion reactor concept" by H. Ruhl and G. Korn (Marvel Fusion, Munich) /1/

K. Lackner

Max Planck Institute for Plasma Physics, 85748 Garching, Germany

Abstract:

The declared aim of Marvel Fusion is the realization of a reactor based on the aneutronic fusion of proton and boron-11 nuclei /2/ in the near future, making use of latest-day laser technology and nano-structured materials. The aim of the preprint quoted above is to demonstrate the feasibility of fusion energy gain from the irradiation of an initially *uncompressed* target consisting of an array of nano-wires, by a femto-second laser. This proposal is in apparent contrast to text-book wisdom, which postulates - even for the fast-ignitor concept, and a DT fuel mix - a target density about 1000 times that of solid state (refs./3/,//4/). The novel, optimistic predictions of Ruhl and Korn are based, however, not on rigorous estimates, but only on parametric dependencies, extrapolated far beyond their conventional limits of validity. The authors invoke the effects of self-arising magnetic fields to result in additional magnetic confinement, but we show their model to contain intrinsic contradictions. The conclusion from out note is that ultra-fast, high power lasers can be of use to inertial fusion energy production only, if the intrinsic time-scale for fusion reactions can be brought down to a level more commensurate with the Laser pulse length. Involving magnetic fields does not change this basic limitation, as the Alfvén time will be similar to the kinetic disintegration time. Given that the fusion energy gain as function of particle energy has a maximum (for DT at about 170 keV) overheating will not be helpful so that the required enhancement of the fusion reaction rate can only be achieved by strong pre-compression of the target.

Introduction:

The declared aim of Marvel Fusion is the realization of a reactor based on the aneutronic fusion of proton and boron 11 nuclei /2/ in the near future, making use of latest-day laser technology and nano-structured materials. The aim of the preprint quoted above is to demonstrate the feasibility of fusion energy gain from the irradiation of an initially *uncompressed* target consisting of an array of nano-wires by a femto-second laser. The novel, optimistic predictions of Ruhl and Korn are based, however, not on rigorous estimates, but only on known parametric dependencies, extrapolated far beyond their conventional limits of validity. They attempt to justify the implied long residence time of energetic ions in the fuel assembly by effects of self-generated magnetic fields.

Estimations of Fusion Gain:

In the two chapters 2 and 3 of their preprint /1/ the authors write down sets of generally valid equations describing the motion and collisional interaction of particles, which are of no further interest for the present discussion as they are not used to determine the estimates in the preprint. Rather, in chapter 4 the authors describe the radial acceleration of ions in the electric field arising after a Coulomb explosion triggered by the interaction of a femto-second Laser pulse with a target consisting of an array of nano-wires. Again, these calculations are of relevance for the estimations in the preprint only by motivating the quantitative choice of an initial energy and radial (in an axisymmetric frame of a nano-wire) ion momentum. Chapter (5) deals with the subsequent slowing down of the ions, where, due to the model chosen (see their equs. (61) and (62)), the radial velocity and the energy of the ions are affected only by drag by the electrons. An axial component of the momentum is supposed to arise because of an azimuthal magnetic field, but its effect on the radial motion is actually ignored in an ordering argument.

This model corresponds to the infinite beam particle confinement model used by Jassby /5/ for the DT reaction, and later extended by, e.g., Rider /6/ to pB and other fusion reactions. Describing only the competition between binary collision effects (electron drag and fusion reactions) the predicted burn-up fraction and energy gain depend practically only on electron temperature and ion starting energy, and only very weakly (through the Coulomb logarithm) on density. For sufficiently high ion energy and electron temperatures /7/, a gain larger than one can readily be achieved for DT. A rudimentary form of energy balance for electrons can easily be added to the model /6/, and - considering only bremsstrahlung losses - a positive energy gain is just marginally possible even for the proton-boron reaction, if the latest adjustments in the fusion cross-section data are taken into account /8/.

These results, and in particular the lack of a dependence on density, however, only hold if no competing ion particle or energy losses occur. The relevance of different loss processes is best highlighted by a comparison of their time scales. For the least demanding DT reaction, and the T density ($n_T = 5 \cdot 10^{28} m^{-3}$), electron ion collisionality ($10^{10} s^{-1}$) and D exit velocity ($v_D = 2 \cdot 10^7 m\ s^{-1}$ - corresponding to an energy $\epsilon_D = 4\ MeV$) quoted as a reference case by the authors, the characteristic time for a D-ion to produce the equivalent of its starting energy by fusion reactions with T atoms

$$\tau_f = \frac{\epsilon_D/\epsilon_f^{DT}}{n_T \sigma_f(\epsilon_D) v_D}$$

is, however, larger than $10^7$ femtoseconds (fs). In a purely radial motion, the particle will leave the laser-illuminated region, however, already in 500 fs! - much too fast for an energetically significant number of reactions to happen. In this model the direct escape of the particles would happen practically unimpeded, as the electron drag is too weak (its characteristic time for the case considered is $10^5$ fs) and also does not change the direction of ions. The requirement to have a sufficient number of fusion reactions before disintegration in fact takes the well-known form of a minimum line-density requirement $\rho R$ ubiquitous in inertial fusion, which can only be met by strong pre-compression. Also, a magnetic field in the approximation introduced into the equations of motion - equs. (61)-(63) - would not affect the radial escape velocity, but only add an axial drift.

While the conversion ratio - the total number of fusion reactions triggered by a fast ion during its slowing down - increases monotonically with starting energy, the energy conversion efficiency - measuring the energy release by the nuclear fusion reactions divided by the kinetic energy of the initiating ion - does not. For DT it has a maximum at around 170keV, and for the parameter combination chosen by RK and used also above it has already decreased by an order of magnitude! Without precompression, and the resulting density increase, the ions are "overheated" by the high-power density of the laser, as noted already by Földes and Pokol /9/ in their comment to other proposals involving nanoplasmonics: "inertial fusion without compression does not work either with or without nanoplasmonics".

Ruhl and Korn specifically mention the possibility to use a range of neutronic and aneutronic fuels, possibly also in combinations. While, e.g., a triggering of DD reactions by a primary stage of DT has computationally been shown to be in principle possible /3/, such options can only work if the most easily initiated reaction (DT) can produce energy amplification.

One alternative formulation of the ignition/breakeven requirement, used e.g., in ref./5/, and customary in stationary magnetic confinement is to lump non-volumetric energy losses into an energy confinement time. In the formulation of RK this implies that the energy loss-rate is not determined by the electron drag as in equ. (71), but by escape of ions out of the system. The quantity $\Delta t$ in their equ. (71) should therefore be viewed, more generally, as a characteristic loss or residence time. A reference value would be given - for the example quoted - by the 500-fs time-of-flight. For a substantial fusion gain, the ions would therefore have to remain trapped in the laser spot region for many thousands of such transits. Confining ions magnetically would appear, at first glance, a possibility to achieve this aim.

It has indeed been shown by Kaymak et al./10/, that the irradiation of nano-wire arrays with sufficiently intense laser-pulses can drive electron currents in the low-density plasma in the gaps, with return currents flowing within the nano-wires. This phenomenon has been well established by code-calculations, but can be also understood semi-quantitatively in terms of an analytic model/11/. The produced magnetic fields are predicted to be very strong (up to megatesla) and to compress the nano-wires. For fusible materials this will cause bursts of nuclear reactions. These phenomena were analyzed and discussed so far on the laser pulse time-scale (some tens to hundreds of fs), but to be energetically relevant would have to persist for a burn-time - i.e., more than $10^6$ fs/12/.

# Improving Fusion Yield by Magnetic Confinement

This requires the establishment of quasistatic magnetic confinement lasting over a fusion-reaction time-scale. The set of equations cited by the authors of /1/ and their solution, however, cannot support this claim, as they are partly inconsistent, or represent an invalid ordering.

The authors first write down the electron part of a two-fluid model: equs. (78) - (83). Their proposed set contains, however, *two* momentum equations (equs. (79) and (82)) for the electron fluid velocity

$$v_e = j_e/(e_e n_e),$$

as being simultaneously valid, which represent in reality two different, mutually contradicting approximations.

Their equation (79), or - if electron inertia (justifiably) is neglected: equ. (84) - implies that the electron $\vec{j}_e \times \vec{B}$ force is balancing the electron pressure, whereas equ. (82), or - equivalently equ. (87) implies neglect of the electron pressure, and balances $\vec{j}_e \times \vec{B}$ only by ion forces transmitted via the electrostatic field. Both situations do occur in magnetic confinement (they are related to the electron and ion root, respectively) but constitute opposite physics regimes for one equation.

The correct electron momentum equation has to contain also the electric field; i.e. (79) has to be extended to read

$$\left(\frac{\partial}{\partial t} + \vec{v}_e \cdot \frac{\partial}{\partial \vec{x}}\right) \vec{v}_e = \frac{1}{m_e n_e} \vec{j}_e \times \vec{B} + e_e \vec{E} - \frac{1}{m_e n_e} \frac{\partial P_e}{\partial \vec{x}},$$

where we have added the index $e$ to $P$ to clarify the use of the electronic pressure by Ruhl and Korn.

In fact, treating both equations (84) and (87) to be valid simultaneously gives rise to nonsensical conclusions, producing formally

$$e_e \vec{E} = -\frac{1}{m_e n_e} \frac{\partial P_e}{\partial \vec{x}}.$$

which would further imply, based on the complete electron momentum equation also $\vec{j}_e \times \vec{B} = 0$, and hence also no electron current, and in this model therefore also no magnetic field.

Fusion reactions are driven, however, by the ions and also the description of ion dynamics in this preprint is inconsistent, as it treats the *dominant* effect as a *perturbation*. The authors include the magnetic field in equ. (89) to produce an axial drift, but neglect it in the radial dynamics (equ. 88), where it would actually avoid the radial escape of the particles and confine the ions. This is not consistent with the postulate (equ.96) that the ions be magnetized, which implies that the radial momentum of individual particles would periodically reverse.

In the chain of arguments of the authors this is used for an estimation of the required magnetic field: however, the requirement (equ. 96) that the ion gyroradius should be smaller than the laser-spot is not sufficient to ensure magnetic confinement. Unless the magnetic energy density is comparable to the kinetic one (in magnetic confinement theory usually expressed as $\beta \approx O(1)$) the magnetic field would just be blown away by the plasma.

The gyromotion cannot be treated as a perturbation if adequate magnetic confinement is to be realized, but is *the* zero-order effect: rather the drifts (both $\vec{E} \times \vec{B}$ and $\nabla |B|$) will constitute a perturbation, and will also make an ionic contribution to the plasma currents. For a self consistent determination of the value of the $\vec{E}$-field, as well as for the splitting of the currents between electrons and ions a two-fluid model treating both species is needed.

In fact, this inconsistency of the model with the assumptions needed to realize magnetic confinement also shows up in the numerical results of the authors (equ. (101)), forcing them to conclude: " implying that the parametric solutions in (93) - (95) have to be improved". - In fact, it is not an option to improve this solution procedure - it has to be inverted. The effects arising here only as perturbation (the gyromotion) have to be treated as zero order.

However, the arguably most important conclusion can be drawn already from a simple one fluid model, lumping together ions and electrons, and thereby eliminating the electric field. The current pattern produced by a laser pulse impinging on a nano-wire produces a Z-pinch like magnetic field configuration, with both the forward and backward currents flowing in the plasma of the nano-wire and its surrounding/10/. The equilibrium in a Z-pinch can be approximated by the so-called Bennet-relation (e.g., pg. 87 of ref./13/):

$$\langle P \rangle = \frac{\mu_o I^2}{8\pi F}$$

linking the cross-section averaged pressure to the total axial current *I* and the cross-section area *F* . Taken at the boundary of the whole plasma, *I* will vanish if the currents close entirely within it, and no stationary magnetic confinement of a finite plasma pressure is possible. This is transparent in such an idealized geometry, but the conclusion is a manifestation of the very general virial theorem, which states that magnetic confinement cannot be achieved without the aid of external currents (e.g., pg. 84 of ref./13/)/14/. In fact, the plasma and magnetic field development in the calculations of e.g., ref. /10/ is highly dynamic, affected, besides by the finite laser pulse length also by the Alfven-dynamics. - Z-pinches are also known to be intrinsically unstable, but this issue would only arise, if an equilibrium were actually formed.

In the absence of an equilibrium, the plasma - consisting of the debris of the nano-wires - will disintegrate on the scale of the Alfven-time. For the magnetic field to have any influence on the plasma, it has to have an energy density comparable to the kinetic one of the plasma (i.e., $\beta \approx O(1)$), which in turn, implies that the Alfven-time will be comparable to the kinetic disintegration time of the plasma: by orders of magnitude too short for useful energy production by fusion reactions even for the most favorable DT reaction.

Inertia alone can of course suffice to confine a plasma for a sufficient time to produce a positive fusion yield. This, however, is expressed by the classical $\varrho R$ constraint (see, e.g., Ref /3/) and requires - for peaceful applications - the known very high precompression of the fuel. We have shown here, that also self-generated magnetic fields will not significantly delay the disintegration of the fuel pellet.

Conclusions:

We conclude that the micro-reactor proposed in this preprint will fall far short of achieving useful energy amplification. This conclusion is, however, not restricted to specific features of this particular proposal. It derives from a basic mis-match: as noted also in ref./9/, the inevitably long time-scales for fusion reactions in an uncompressed medium do not allow them to benefit from the capabilities of extremely high-power lasers. Note in particular that overheating the ions with extremely high power does no improve fusion gain: e.g., for DT the maximum energy amplification by fusion reactions is achieved for single particle energies around 170 keV/15/, and decreases by about an order of magnitude already for the ion energies (4MeV) considered by Ruhl and Korn.

The compression need for the fast ignitor is determined by the need to bring the fusion time-scale sufficiently below the thermal disintegration time (other requirements, like the need for fusion product redeposition, or of a limit to electron heat conduction losses in the flux limited regime give similar parametric dependences). Involving magnetic fields does not change this basic limitation, as the Alfvén time will be similar to the thermal transit time. The Laser heat pulse (or the energetic particle pulse produced by it) initiating burn has then just to be short compared to this time-scale so that the energy deposited by it will be available during the whole burn. This holds already for DT fusion concepts, but would become much more demanding for pB fusion.

The absolute need for strong pre-compression for a fast ignitor would seem to limit also the usefulness of nano-structures, which appear difficult to reconcile with it, unless protected by some form of cone-structures like described in, e.g., refs. /3/, /4/.

References

[1] H. Ruhl and G. Korn: A laser-driven mixed fuel nuclear fusion reactor concept, arXiv:2202.03170v4 [physics.plasm-ph] 8 Mar 2022

[2] https://marvelfusion.com/technology/

[3] S. Atzeni, J. Meyer-ter-Vehn: The Physics of Inertial Fusion, Oxford Science Publications, 2004

[4] K. Mima: Inertial Fusion Energy, in M. Kikuchi, K. Lackner, M.Q. Tran (editors): Fusion Physics, IAEA Vienna, 2012

[5] D.L. Jassby: Neutral-beam-driven tokamak fusion reactors, chapt. 6, Nucl. Fusion 17 309 (1977)

[6] T.H. Rider: Fundamental limitations on plasma fusion systems not in thermodynamic equilibrium, Physics of Plasmas 4, 1039 (1997)

[7] J.M. Dawson et al.: Production of Thermonuclear Power by Non-Maxwellian Ions in a Closed Magnetic Field Configuration, Phys. Rev. Lett. 26, 1156 (1971)

[8] S.V. Putvinski et al.: Fusion reactivity of the pB11 plasma revisited Nucl. Fusion 59 076018 (2019)

[9] I.B. Földes and G.I. Pokol: Inertial fusion without compression does not work wither with or without nanoplasmonics, Laser and Particle Beams 38, 211 (2020)

[10] V. Kaymak, A. Pukhov, V.N. Shlyaptsev and J.J. Rocca: Nano-scale ultradense Z-Pinch Formation from Laser-Irradiated Nanowire Arrays, Phys. Rev. Lett. 117,035004 (2016)